\documentstyle[12pt]{article}
\begin{document}
\centerline{\bf Low-lying excitations in superconducting bilayer systems} 
\baselineskip=22pt
\vspace{2cm}
\centerline{ Marius Grigorescu } 
\centerline{ Department of Physics and Astronomy}
\baselineskip=12pt
\centerline{ University of Western Ontario}
\centerline{ London, Ontario, Canada N6A 3K7}
\vspace{5cm}
{\bf Abstract}: \\
The ground and first excited state of two superconducting layers in 
interaction are studied considering two different coupling terms, 
one represented by the standard Josephson interaction, and one new, which 
is a superexchange pairing force between bilayer pairs. It is shown that a 
moderate-to strong Josephson interaction produces a low-lying collective 
state, pictured as an out-of-phase oscillation of the BCS gauge angles of the 
two layers. 
This antisymmetric angular oscillation might explain the 41 meV resonance 
observed in the neutron scattering experiments. The bilayer pairs are formed 
by electrons from different layers with an antiparallel orientation of the 
spins, being related to the antiferromagnetic arrangement. The pair 
operators within the layers together with the bilayer pairs generate by 
commutation an so(5) algebra. It is shown that the transition between the 
superconducting and antiferromagnetic phases can be explained assuming
the dependence on concentration of the bilayer pairing strength, with 
maximum at half-filling.
\\[1cm]
PACS numbers: 71.10.Li, 74.50.+r, 74.72.-h, 03.65.Fd   
\newpage
{\bf I. Introduction} \\[.5cm] \indent
Since the discovery of the high-T$_c$ superconductivity \cite{bm} the
remarkable properties of the cuprate compounds have been a subject of
thorough experimental and theoretical investigation. Although a generally 
accepted theory is not yet established, the extensive knowledges accumulated 
during the years indicate that there are few key elements closely 
related to the mechanism of the high-T$_c$ superconductivity \cite{and}. 
\\ \indent
A common element in all cuprate compounds is the layered structure,
consisting of CuO planes separated by block layers \cite{imf}. There 
is also a single parameter, the occupation fraction, whose variation may 
switch the state between superconducting (SC) and antiferromagnetic (AF). For 
example, when the planar Cu atoms of La$_2$CuO$_4$ or YBa$_2$Cu$_3$O$_6$ are 
divalent each site contains one particle (half-filling), and the compounds 
are AF insulators. However, if an increasing  number of Cu atoms become 
trivalent, (hole doping), the AF order is destroyed, and after a transitory 
spin-glass state the compounds becomes metallic, or below T$_c$, 
superconducting \cite{scal}. 
\\ \indent
Although the pair binding energy is supposed to come mainly from the 
interactions within the planes, the interlayer coupling is particularly
important. Thus, the properties of the high-T$_c$ compounds become more 
isotropic in the SC state \cite{and}, and the measurements indicate that the 
current normal to the planes is of Josephson type \cite{fish}. In the 
insulating phase, the AF ordering is not restricted to the nearest neighbors 
within the planes. It appears also between the adjacent layers \cite{af1, af2}, 
and is due to a weak AF interlayer exchange interaction  \cite{af3, ss}. 
\\ \indent
An interlayer coupling of Josephson type can amplify the pairing correlation
within the CuO planes, explaining the high T$_c$-values and the 
condensation energy \cite{csas, and1}. With this coupling the layered 
cuprates resemble a stack of Josephson junctions  \cite{tink}, each of 
them being represented by two adjacent layers (bilayer) coupled by 
the Josephson interaction.   \\ \indent
The low-lying spectrum of the bilayer system is of particular interest, and 
when the layers become superfluid one can expect collective excitations 
related to the phase of the BCS (Bardeen-Cooper-Schreiffer) order parameter  
\cite{sch, and1}.  
As it was shown for the first time in \cite{mg}, a time-dependent mean-field
treatment (TDMF) of the Josephson coupling in a two-component SC system 
indicates the occurrence of a low-lying resonance. This is antisymmetric 
between the two components, and can be pictured as an oscillation of the 
relative phase angle between the two BCS condensates. However, the TDMF 
approximation is quasi-classical, and therefore in this approach purely 
quantum effects as tunneling and quantum coherence oscillations are usually 
suppressed \cite{mg1}. 
\\ \indent
The relevance of the TDMF phase angle resonance for the low-lying spectrum of 
a SC bilayer system will be studied in Sect. I. 
Each component of the bilayer will be described by a reduced BCS interaction
term, and spectrum is calculated numerically as a function of the strength 
of the Josephson coupling and the occupation fraction. It will be shown that 
for moderate-to-strong coupling the phase angle resonance gives a very
good description of the first excited state. This resonance is 
antisymmetric between the planes, appears only in the superconducting state, 
is coherent between the CuO planes and of nonmagnetic origin. By these 
properties it can provide an explanation for the 41 meV resonance observed 
in the neutron scattering experiments on YBa$_2$Cu$_3$O$_7$ \cite{nse1,nse2}.  
\\ \indent
If the two layers can exchange particles, then a pair of
electrons localized in one plane may be changed into a pair of 
electrons from different planes. Such "bilayer pairs" appear also when
the Josephson coupling term between the two layers is derived from a
general two-body pairing interaction. Because these pairs contain electrons 
from different planes with an antiparallel arrangement of the spins, 
they can be naturally related to the AF inter-layer ordering.  Following the 
example presented in \cite{mg}, the extension of the su(2)$\oplus$su(2) 
algebra of the BCS pairing operators within the planes by the bilayer pair 
operators leads to the so(5) algebra, and it will be presented in Sect. III.
\\ \indent
A model of the high-T$_c$ superconductors where so(5) is generated by the 
three operators of the total spin, the particle number operator, and six pair 
operators which generate "rotations" between the total SC order parameter and 
the N\'eel vector was formulated in \cite{scz, rkdz}. The AF state at 
half-filling appears as a solid formed by Cooper pairs, and AF-SC transition 
corresponds to the melting of this solid under the influence of a fictious 
"B-field", represented by the chemical potential \cite{scz}. 
However, in this approach the separation of the system in the 
superconducting components associated to the CuO planes is not explicit, 
and it is difficult to establish the connection between the model parameters 
and the strength of the interlayer Josephson coupling. Therefore, 
the 41 meV excitation observed in YBa$_2$Cu$_3$O$_7$ appears as a 
Goldstone mode associated to the breaking of the SO(5) symmetry rather
than to the "restoring force" determined by the Josephson interaction, and 
its generators have both SC and AF components \cite{hen}. 
\\ \indent
The pairing interaction between the bilayer pairs considered here resemble 
the superexchange interlayer interaction \cite{and}, and is 
complementary to the Josephson coupling. The properties of the low-lying 
spectrum of the  bilayer system will be studied as a function of the strength 
of this interaction and the occupation fraction in Sect. IV. It will be 
shown that by contrast to the Josephson coupling, the new interaction may 
suppress the superconductivity of the planes, leading to an AF arrangement.  
A summary of this study and the conclusions are presented in Sect. V.
\newpage 
{\bf II. Low-lying excitations induced by the Josephson coupling } \\[.5cm] 
\indent
The model Hamiltonian considered here is supposed to describe the electrons 
of two close CuO layers (bilayer system), denoted $A$ and $B$. Assuming that 
these layers are related by a translation along the vector $Q$, between their 
lattice sites there is a one-to-one correspondence expressed by $b=a+Q$. 
The electrons of each CuO plane are supposed to interact strongly by pairing 
forces, and within a restricted BCS approximation, the Hamiltonian has the
form
\begin{equation}
H_{0 L} = \epsilon N_L -G_L P^\dagger_L P_L~~,~~L=A,B.
\end{equation}
Here 
\begin{equation}
N_A= \sum_a ( c^\dagger_{a \uparrow} c_{a \uparrow} + 
c^\dagger_{a \downarrow} c_{a \downarrow} ) ~~,
N_B= \sum_a( c^\dagger_{a+Q \uparrow} c_{a+Q  \uparrow}
+c^\dagger_{a+Q \downarrow} c_{a+Q \downarrow} ) ~~,~~
\end{equation}
are the particle number operators, 
\begin{equation}
P^\dagger_A = \sum_{a, a'} \Phi (a-a') c^\dagger_{a \uparrow}
c^\dagger_{a' \downarrow}~~,~~    
P^\dagger_B = \sum_{a, a'} \Phi (a-a') c^\dagger_{a+Q \uparrow}
c^\dagger_{a'+Q \downarrow}~~,
\end{equation}
create electron pairs with the wave function 
$\Phi(a-a')= \Phi(a'-a)$, and  $c^\dagger_{a m}$ ($c_{a m}$) are the 
creation (annihilation) operators for an electron at the site $a$ ($a+Q$) 
of the lattice $A$ ($B$) with the Z-component of the spin up ($m= \uparrow$) 
or down ($m= \downarrow$). The layers are coupled by a Josephson interaction
term \cite{csas}
\begin{equation}
H_J =  -G_J ( P^\dagger_A P_B + P^\dagger_B P_A) ~~,
\end{equation}
such that the total Hamiltonian has the form
\begin{equation}
H = \epsilon_A N_A + \epsilon_B N_B - G_A P^\dagger_A P_A -
G_B P^\dagger_B P_B - G_J ( P^\dagger_A P_B  + P^\dagger_B P_A)~~.
\end{equation}
The model parameters are  the single-particle energies $\epsilon_A$, 
$\epsilon_B$, the strengths $G_A$, $G_B$, $G_J$ of the pairing 
forces and the number of the lattice sites.
\\ \indent
The commutation relations  between the pair creation operators and 
their Hermitian conjugates are determined by the orbital wave
function $\Phi(a-a')$.  For a $d_{x^2-y^2}$-pair field, this function 
can be expressed as \cite{scal}
\begin{equation}
\Phi(a-a') = \frac{1}{2}( \delta_{a-a', x} + \delta_{a-a',-x} 
 - \delta_{a-a',y} - \delta_{a-a',-y} ) 
\end{equation}
where $x,y$ are vectors along the x and y axes of the plane lattice 
relating the nearest neighbors. The commutator $[P^\dagger_A, P_A]$ 
contains the sum 
\begin{equation}
\sum_{a_1} \Phi(a-a_1) \Phi(a_1 -a') = \delta_{a,a'} +
\frac{1}{4} ( \delta_{a-a',2x} + \delta_{a-a',-2x} + \delta_{a-a',2y} 
\end{equation}
$$
+ \delta_{a-a',-2y} - 2 \delta_{a-a',x+y} - 
2 \delta_{a-a',-x-y} -2 \delta_{a-a',x-y} - 2 \delta_{a-a',-x+y} ) ~~.
$$
Here only the first term of the right-hand side, which is diagonal, relates 
nearest neighbors.  Assuming that the contribution due to the other terms  
can be neglected, and that the number of lattice sites $\Omega$ is the same 
for $A$ and $B$, then $[P^\dagger_L, P_L] = N_L- \Omega$, $L=A,B$. 
Defining $P_{0L}= (N_L- \Omega)/2$,  then $[P_{0L}, P^\dagger_L] = 
P^\dagger_L$, and each set of three operators $(P^\dagger_L, P_{0L}, P_L)$, 
$L=A,B$, generates an su(2) algebra.  
\\ \indent
If $G_J=0$ the ground state $\vert g (N_p)>$ of $H$ is non-degenerate when 
the total number of pairs $N_p$ is even, and double degenerate when it is odd. 
Thus, $\vert g (2 k) >$ can be factorized in a product of two $k$-pair states, 
one for each layer, as $\vert g (2k)>= \vert A(k)> \vert B(k) > $, where  
\begin{equation}
\vert L(k) >= \sqrt{ \frac{ k! \Omega ! }{ ( \Omega -k)!} } (P_L^\dagger)^k 
\vert 0> ~~,~~L=A,B 
\end{equation}
and $\vert 0>$ is the particle vacuum. When $N_p=2k+1$ the ground-state 
doublet states are  $\vert g_a (2k+1) > 
= \vert A(k+1) B(k)>$ and  $\vert g_b (2k+1)> = \vert A(k) B(k+1)>$. 
The term $H_J$ rise the degeneracy between these states, leading to the 
entanglement of the two planes.  When  $G_J$ is small, the 
ground and first excited state of the coupled system are 
\begin{equation}
\vert g_w (2k+1)  > = \frac{1}{ \sqrt{2}}( 
\vert A(k+1)  B(k) >+ \vert A(k) B(k+1)> ) ~~,~~ 
\end{equation}
and
\begin{equation}
\vert 1_w > =  \frac{1}{ \sqrt{2}}( 
\vert A(k+1)  B(k) >- \vert A(k) B(k+1)> ) ~~.
\end{equation}
The difference between their energies $E_x=E_1-E_0$ is given in this case by 
the  tunnel splitting 
$ E_t= 2 \vert < A(k+1) B(k) \vert H_J \vert A(k) B(k+1)> \vert =
2 G_J \Omega^2 (n+  \Omega /2 )(1-n+\Omega /2 ) $,
where $n=N_p/2 \Omega$ denotes the occupation fraction.    
\\ \indent
This result was compared with the excitation energy of the first level
calculated numerically for a bilayer system with $\Omega=12$, $\epsilon_A=
\epsilon_B=0$ and $G_A=G_B=1$, as a function of $n$ and $G_J$. The first 
excited state is always antisymmetric between the planes, and when $G_J$ is 
smaller than $\sim 0.05$, $E_x$  shows a strong odd-even effect at the 
variation of $N_p$. This behavior is illustrated for 
$G_J = 0.005$ in Fig. 1 (a) (dots joined by dashed lines). When 
$G_J > 0.1$, this strong odd-even effect disappears, as it is shown 
in Fig. 1 (c) for $G_J = 0.2$.  The comparison with $E_t$ (Fig. 1 (a,c), 
solid line) shows a good agreement when $N_p$ is odd and $G_J$ is small, 
but for $N_p$ even or when $G_J$ increase the differences become large,
especially at half-filling. This result indicates that at moderate-to-strong 
coupling ($G_J > 0.1$) the tunneling approach to the Josephson 
interaction is not realistic. It is interesting to note that the range of the 
$G_J$ values when the coupling is "weak" or "strong"  depends of $\Omega$. 
For instance, decreasing  $\Omega$ from 12 to 6, the interval separating the 
weak and strong coupling regimes changes from $\sim (0.01,0.1)$ to 
$\sim (0.05,0.2)$.  
\\ \indent
The pair transfer matrix element between the ground states,  
$\Delta = G < g (N_p) \vert P^\dagger_L \vert g (N_p-1) >$ is a measure of 
the "softness" of the system with respect to the variation of the number of
pairs. A high value of $\Delta$ indicates the increase of the two-particle
correlations in the ground state, and may signal the transition to a 
superconducting state. This parameter was calculated for $G_J=0.005$ and 
$0.2$, and the results are represented in Fig. 1 (b) and (d), respectively.
Their comparison confirms the increase of the pairing correlations by 
the interlayer Josephson interaction, and shows that this effect is stronger 
at half-filling. \\ \indent
When a layer becomes superconducting, its ground state breaks the U(1) 
symmetry of the Hamiltonian generated by the particle-number (charge) 
operator, and is well approximated by the  BCS function
\begin{equation}
\vert BCS ( \phi, \lambda) > = e^{( 2 z P^\dagger_L - 2 z^* P_L)} \vert 0> ~~,
L=A,B~~.
\end{equation}
Here $\lambda$ is the Fermi energy, $\phi$ is the BCS "gauge angle", 
$z= \rho e^{-i \phi}$, and $\rho$ is a variable related to the BCS gap 
$
\Delta_0 = G <BCS (0, \lambda)  \vert P^\dagger_L \vert BCS (0, \lambda)>
$
by $\Delta_0= (G \Omega /2) \sin 4 \rho$, and to the occupation fraction 
$n= <BCS(0, \lambda ) \vert$ $ N_L$ $ \vert BCS(0, \lambda) >/ 2 \Omega $ by 
$n= \sin^2 2 \rho$.  The spontaneous breaking of the U(1) symmetry 
generates a specific collective phase-space \cite{mg94} parameterized by the 
BCS "gauge" angle $\phi$ and the canonically conjugate momentum
$p= \Omega n= \Omega/2  - ( \epsilon - \lambda)/G $ \cite{mg}. 
In the absence of the interlayer coupling, $\phi$ has a linear dependence on 
time, $\phi(t)= 2 \lambda t$, such that the system performs a free rotation 
in the BCS gauge-space, with the angular velocity $2 \lambda$. This 
rotation can be interpreted as the Goldstone mode related to the U(1)
symmetry breaking, and the classical Hamilton function associated to this 
evolution is defined by the expectation value of $H_{0 L}$,
\cite{mg}
\begin{equation}
{\cal H}_0 ( \phi, p) = < BCS \vert H_{0 L} \vert BCS> = 
2 (\epsilon-G \frac{ \Omega}{2} ) p +G p^2~~. 
\end{equation}
This expression indicates that the inertial parameter corresponding to the 
free gauge rotation is $I_0 = 2/ G$. \\ \indent
When the two superfluid layers are coupled by the Josephson interaction term
$H_J$, the Hamilton function becomes 
\begin{equation}
{\cal H}( \phi_A, p_A, \phi_B, p_B) = 
< BCS_A BCS_B \vert H_{0A} + H_{0B} + H_J \vert BCS_A BCS_B > = 
\end{equation}
$$
2 (\epsilon-G \frac{ \Omega}{2}) (p_A+p_B) +G (p^2_A +p^2_B) 
- 2 G_J \sqrt{ p_A p_B ( \Omega -p_A) ( \Omega -p_B) } \cos ( \phi_A -
\phi_B)~~. 
$$
In this case, the dynamics of the system consists of a superposition between 
a uniform rotation of the "center of mass" angle $(\phi_A + \phi_B)/2$, 
symmetric between the planes, and an oscillation in the relative angle 
$(\phi_A - \phi_B)/2$, which is antisymmetric. Assuming the same occupation 
fraction $n$ for both planes, in the harmonic approximation the excitation 
energy of this mode is \cite{mg} 
\begin{equation}
E_r =  \sqrt{2 k C} ~~.
\end{equation}
Here $k=2(G-G_J)+ G_J/n(1-n) $ is the related inertial parameter, and 
$C= 2 G_J ( \Delta_0/ G)^2= 2 G_J \Omega^2 n (1-n) \approx E_t$, 
is the strength of the restoring force  determined by the Josephson coupling.  
The same excitation energy is obtained when $H_J$ is treated within the 
random-phase approximation (RPA) \cite{gi, mg2}. 
\\ \indent
The energy $E_r$ of the antisymmetric resonance oscillation and the 
BCS gap $\Delta_0$ obtained with the same parameters as used in the numerical 
calculations are represented by solid lines in Fig. 1 (a,c)  and (b,d),
respectively. The agreement with the numerical values is very good at 
moderate to-strong coupling, when $E_x \approx E_r$ and $\Delta \approx 
\Delta_0$. At weak coupling $E_r$ still reproduces the average trend of 
$E_x$, but $\Delta$  is systematically smaller than $\Delta_0$ by a factor 
$\sim 0.83 $. Both $\Delta$ and $\Delta_0$ measure the pair correlation 
functions within a single plane. If each plane develops its superconductivity 
independently, then in the limit $G_J \rightarrow 0$  one would expect 
that $\Delta \rightarrow \Delta_0$. However, this is not observed because 
when $G_J$ is very small, at $N_p$ odd the coupling between the two planes 
leads invariably to an entangled ground state of the form $\vert g_w>$. 
Therefore, the one-pair transfer matrix element $\Delta$ is reduced to 
$1/ \sqrt{2}$ of the value corresponding to an isolated plane. By contrast, 
at moderate-to-strong coupling the ground state is spread over many 
configurations, and BCS approximation becomes realistic.  This shows that 
the interlayer Josephson coupling has an important contribution to the 
occurrence of the superconductivity within the planes. 
\\ \indent
Beside the "kinetic" term $G(p_A^2+p_B^2)$ determined by the pairing 
forces, an additional contribution to ${\cal H}$ 
appears from the electrostatic energy between the planes,
$E_{es} =e^2 (p_A - p_B)^2 d / (2 \epsilon_0 S) $   ( \cite{and}, p. 124), 
where $d$ is the interlayer spacing and $S$ the layer surface. With this
term $E_r$  becomes $\sqrt{2(k+k_e)C}$, where $k_e=2 e^2 d/( \epsilon_0 S)$. 
However, by contrast to $k$, which is size-independent, $k_e$ vanishes at 
the macroscopic limit $S \rightarrow \infty$, and it will be neglected. 
\\ \indent
The condensation energy and the superconducting energy gap are known 
from the measurements of the electronic specific heat \cite{loram1, loram2}. 
For  YBa$_2$Cu$_3$O$_7$ the zero temperature superconducting gap 
has the maximum $\Delta_0 = 20$ meV, and the condensation energy is
$U_c = 0.5$ meV/f.u. cell. Assuming that $\Delta_0 = G \Omega \sqrt{n(1-n)}$, 
with $n \sim 1/2$, and  $G_J/G  \approx 0.5 $, one obtains
$E_r \approx 50$ meV. This low value suggests that the antisymmetric 
angular oscillations in the BCS gauge space presented here may provide an 
explanation for the 41 meV resonance observed in the neutron scattering 
experiments \cite{nse1, nse2}.  
\\ \indent
The general form of the operator creating a pair of electrons with a
three-dimensional orbital wave function $\Psi(l-l')$ in the bilayer system is
\begin{equation}
P^\dagger = \sum_{l, l'} \Psi (l-l') c^\dagger_{l \uparrow}
c^\dagger_{l' \downarrow}~~,
\end{equation}
where $l$  runs over the lattice sites of both planes, $A$ and $B$. 
Assuming that the in-plane section $\Psi(a-a')$ is the same as  $\Phi(a-a')$,  
and $\Psi(l)$ is periodic in the $Q$-direction, $\Psi(l)=\Psi(l+Q)$, this 
general pair creation  operator can  be decomposed as
\begin{equation}
P^\dagger = P^\dagger_A + P^\dagger_B + R^\dagger
\end{equation}
where $P^\dagger_A$ and $P^\dagger_B$ are the pair creation operators  
within the planes of Eq. (3), and 
\begin{equation}
R^\dagger = \sum_{a, a'} \Phi (a'-a) ( c^\dagger_{a+Q \uparrow} 
c^\dagger_{a' \downarrow} + 
c^\dagger_{a+Q \downarrow} c^\dagger_{a' \uparrow} )
\end{equation}
creates a pair of electrons localized in different planes, with an 
antiparallel spin orientation.  \\ \indent
The antiparallel orientation of the nearest spins from different planes is a 
characteristic of the interlayer antiferromagnetic ordering. In fact, 
the AF ordering of the high-T$_c$ materials was described 
within the resonance valence bond (RVB) theory as a condensate of nearest 
neighbors singlet pairs \cite{rvb}. A general pairing interaction of the form
$-G P^\dagger P$ leads naturally the terms of the Hamiltonian $H$, but 
contains also other terms dues to the "bilayer pairing" operators $R$. In
particular, a pairing interaction $- R^\dagger R$ between the bilayer pairs 
can represent a convenient approximation for the superexchange interlayer 
coupling presented in \cite{and}, p. 395. This term can produce the AF 
condensation, and it will be studied by extending to so(5) the 
su(2)$\oplus$su(2) algebra used to describe the effect of the interlayer 
Josephson coupling. 
\\[.5cm]
{\bf III. The so(5) algebra of the pair operators in the bilayer system}  
\\[.5cm]
The shift along the chain
\begin{equation}
P^\dagger_A \rightarrow R^\dagger \rightarrow P^\dagger_B
\end{equation}
of the three pair creation operators defined above is generated by 
commutation with the interlayer "hopping" operator
\begin{equation}
T_+ = \sum_a ( c^\dagger_{a+Q \downarrow } c_{a \downarrow}  -
c^\dagger_{a+Q \uparrow} c_{a \uparrow} )~~,
\end{equation}
such that
\begin{equation}
[T_+, P^\dagger_A] = - R^\dagger ~~,~~ [T_+, R^\dagger ] =
2 P^\dagger_B ~~,~~[T_+, P^\dagger_B]=0 ~~.   
\end{equation}
Similarly, the down shift 
\begin{equation}
P^\dagger_B \rightarrow R^\dagger \rightarrow P^\dagger_A
\end{equation}
is generated by commutation with $T_- \equiv (T_+)^\dagger$. Each set of 
three operators, $(R^\dagger, R_0, R)$ and $(T_+, T_0, T_-)$, with  
$R_0= P_{0A}+P_{0B}$ and $T_0= (N_B-N_A)/2$ generate a new su(2) algebra, 
while the set of the 10 operators $T_+, T_-$, $P^\dagger_A, P_A$, 
$R^\dagger, R$, $P^\dagger_B, P_B$, $P_{0A}, P_{0B}$ generate the so(5) 
algebra.   \\ \indent
It is important to note that by contrast to $P^\dagger_A$ and $P^\dagger_B$ 
which create spin singlet pairs, the pair created by $R^\dagger$ 
is the $m=0$ component of a spin triplet. A spin singlet pair of
electrons from different planes is generated by 
\begin{equation}
S^\dagger = \sum_{a, a'} \Phi (a-a') ( 
c^\dagger_{a+Q \uparrow} c^\dagger_{a' \downarrow} -
c^\dagger_{a+Q \downarrow} c^\dagger_{a' \uparrow} )~~.
\end{equation}
Pairs of this type have been used to construct the RVB vacuum in
\cite{eder}. The shift along the chain
\begin{equation}
P^\dagger_A \rightarrow S^\dagger \rightarrow P^\dagger_B
\end{equation}
is generated by the commutation with the hopping operator
\begin{equation}
\tau_+ = \sum_{a}( c^\dagger_{a+Q \downarrow } c_{a \downarrow}  + 
c^\dagger_{a+Q \uparrow} c_{a \uparrow} )~~,
\end{equation}
and  $\tau_+, \tau_-$, $P^\dagger_A, P_A$ , $S^\dagger, S$, 
$P^\dagger_B, P_B$, $P_{0A}, P_{0B}$ generate also an so(5) algebra.   
\\ \indent
In the defining representation, the so(5) generators are
$5 \times 5$ matrices $f_{p q}=e_{pq}-e_{-q -p}$, where the 
indices $p$ and $q$ take the values -2,-1,0,1,2, and $e_{pq}$ is the 
$5 \times 5$ matrix with the element on the row $p$ and the column $q$ 
equal to 1, and all the rest 0.  Thus 
$[e_{pq},e_{kl}]= \delta_{qk} e_{pl} - \delta_{lp} e_{kq}$, and  the 
commutation relations of the so(5) generators are 
\begin{equation}
[ f_{pq}, f_{kl}] = \delta_{qk} f_{pl} - \delta_{pl} f_{kq} +
\delta_{p -k} f_{-l q} - \delta_{- q l} f_{p -k}~~.
\end{equation}
The correspondence between the generators with negative roots of the 
defining representation of so(5) and the two different realizations provided 
by the pairing operators, denoted so(5)$_r$ and so(5)$_s$, is summarized in 
the Table 1. 
One should note that these two realizations are related simply
by a change of sign for all the single-particle states with spin up in the 
layer $A$. This transformation  can be written as
$c^\dagger_{a \uparrow} \rightarrow - c^\dagger_{a \uparrow}$, and as long 
as it leaves the Hamiltonian invariant, the results are independent on
the choice of so(5)$_r$ or so(5)$_s$ for the explicit realization. 
\\ \indent
The Cartan subalgebra of so(5) is generated by two elements, 
$f_{22}, f_{11}$, and in both realizations 
\begin{equation}
f_{22} = \frac{1}{2} (P_{0B} + P_{0A})~~,~~ f_{11}= \frac{1}{2} ( 
P_{0B}-P_{0A})~~.
\end{equation}
The algebra  so(5) is semisimple of rank 2, and  its
irreducible representations can be labeled by the numbers
$[N_1, N_2]$, $N_1 \ge N_2 \ge 0$ of the associated Young diagram, or by 
the eigenvalues of the generators of the Cartan subalgebra for the highest 
weight state \cite{rowe}. If this state is chosen to be the particle 
vacuum, $\vert 0>$, then similarly to the case of su(3) \cite{mg84}, the 
states $\vert k l m>$ 
\begin{equation}
\vert k l m  > = 
(f_{2 -1})^k (f_{12} )^l (f_{02} )^m \vert 0 >  ~~,
\end{equation}
with $k+l+m \le 2 \Omega$ span the Hilbert space ${\cal H}_\Omega$, carrier 
of the symmetric irreducible representation $[ \Omega, 0]$ of so(5). 
\\ \indent
The labeling of the states of this irreducible representation requires
three indices. One of them is provided by the total number of
particles $2(k+l+m)$, which takes values between 0 and $4 \Omega$,  
and another by the eigenvalue $k-l$ of $T_0$ ($=\tau_0$). The choice of the 
third index is a difficult long-standing problem, because the set of all 
possible states $\vert k l m>$ with a fixed $N$ and $k-l$ is non-orthogonal 
and overcomplete \cite{hecht}. For instance,  there is a non-vanishing 
amplitude  $<k k 0 \vert 0 0 2k> = (2k)! k! \Omega! /(\Omega-k)!$ 
for recoupling $2k$ bilayer pairs in two sets of $k$ one-layer pairs.  
\\ \indent
The basis of the $[ \Omega, 0]$ irrep spaces can be obtained by using 
vector coherent states methods, by projection from a highest weight state 
\cite{rh}. Here the basis will be constructed numerically, restricting the 
overcomplete set of Eq. (27) by the Gramm-Schmidt procedure. As a test, the 
dimension of the spaces ${\cal H}_\Omega$ generated for $2 \le \Omega \le 12$ 
was found to be the same as given by the analytical formula
\begin{equation}
d(\Omega, 0) = \frac{1}{6} ( 2 \Omega +3) (\Omega +2) ( \Omega +1)~~.
\end{equation}
The action of the generators on the (non-orthonormal) states of Eq. (27) is 
expressed by
$ f_{2 -1} \vert k l m> = \vert k+1 l m> $, 
$ f_{21}  \vert k l m> = \vert k l+1 m> $, 
$ f_{20} \vert k l m> = \vert k l m+1>  $,
and 
\begin{equation}
f_{-12} \vert k l m> = -k (k+m- \Omega -1) \vert k-1 l m>
+ m(m-1) \vert k l+1 m-2>
\end{equation}
\begin{equation}
f_{12}  \vert k l m> = -l (l+m- \Omega -1) \vert k l-1 m>
+ m(m-1) \vert k+1 l m-2> 
\end{equation}
\begin{equation}
f_{02} \vert k l m> = kl \vert k-1 l-1 m+1>
-2 m(k+l- \Omega+\frac{m-1}{2} ) \vert k l m-1> .
\end{equation}
These formulas are useful to calculate the matrix elements of the general
pairing Hamiltonian in the so(5) basis. 
\\[.5cm]
{\bf IV. SC-AF transition induced by the bilayer pair interaction } 
\\[0.5cm] \indent
The extended Hamiltonian which includes the interaction between the
bilayer pairs has the general form  
\begin{equation}
H_g =  - G (P^\dagger_A P_A + P^\dagger_B P_B ) - 
G_J ( P^\dagger_A P_B  + P^\dagger_B P_A) - g R^\dagger R ~~.
\end{equation}
This Hamiltonian is expressed only in terms of the
so(5)$_r$ generators, and therefore it allows to restrict the Hilbert space  
to a space ${\cal H}_\Omega$ of irreducible representation. 
\\ \indent
The low-lying spectrum of $H_g$ has been calculated for $G=1$, 
$G_J=0.5$ and $\Omega =6$, as a function of the strength $g$ of the 
bilayer pairing interaction and the occupation fraction $n$.  The ground 
state energy $E_0$ and the excitation energy of the first level 
$E_x=E_1-E_0$ are represented as a function of $n$ and $g$ in Fig. 2(a) and 
2(b). The pair transfer matrix element $\Delta$ is represented in Fig. 3(a). 
\\ \indent
For the study of the AF correlations each lattice $L=A,B$ is separated 
in two sublattices, denoted by $(L,1)$ and $(L,2)$, such that the 
nearest neighbors of the spins from $(L,1)$ within the plane $L$ are in 
$(L,2)$. If the nearest neighbors of $(A,1)$, $(A,2)$ outside the $A$-plane 
are in $(B,1)$, $(B,2)$, then in a classical antiferromagnetic configuration 
the spins from $(A,1)$ and  $(B,2)$ have all the same orientation, opposite 
to that of the spins from $(A,2)$ and $(B,1)$. \\ \indent
In a quantum system the situation is more complex, because the ground state
is a symmetric superposition of classical AF states with opposite orientations 
of the N\'eel vector \cite{mgms}.  Thus, as a measure of the AF correlations 
in the quantum bilayer system here will be considered the ground state 
expectation value of the operator
${\cal C} = (S_z^A(1)-S_z^A(2))(S_z^B(1)-S_z^B(2))$, where
\begin{equation}
S_z^A ( \alpha) = \frac{1}{2} \sum_{a \in (A, \alpha) }
( c^\dagger_{a \uparrow} c_{a  \uparrow} - 
c^\dagger_{a \downarrow} c_{a \downarrow}) ~~,
\end{equation}
and
\begin{equation}
S_z^B (\alpha)  = \frac{1}{2} \sum_{a \in (B, \alpha) }
( c^\dagger_{a+Q \uparrow} c_{a+Q  \uparrow} -
c^\dagger_{a+Q \downarrow} c_{a+Q \downarrow}) ~~,
\end{equation}
for $\alpha =1,2$ are the z (c-axis)-components of the total spin for each 
sublattice. If only the terms containing nearest neighbors are retained,
then the ground-state expectation value $< g \vert {\cal C} \vert g>$ 
can be approximated by $< g \vert S_z^A(1)S_z^B(1)+S_z^A(2)S_z^B(2) \vert g>$ 
and also by $<g \vert S_z^A S_z^B \vert g>$, where $S_z^L = S_z^L(1)+S_z^L(2)$. 
This approximation is particularly convenient if the Hilbert space
is restricted to ${\cal H}_\Omega$. In this case, the 
matrix elements of $S_z^A S_z^B$ can be calculated analytically, and have
the expression
\begin{equation}
< k' l' m' \vert S_z^A S_z^B \vert k l m> = 
\lbrack (k+ \frac{m}{2})(l+ \frac{m}{2}) - kl - m \frac{ k +l +m}{2} +
\end{equation}
$$
\frac{ m(m-1)}{2} \frac{( \Omega^2 -1)}{\Omega (2 \Omega -1) -1} 
\rbrack f  + \frac{ m(m-1)( \Omega -1)}{ \Omega ( 2 \Omega -1) -1} g ~~,
$$
where
$f=< k' l' m' \vert k l m>$ and  $g=< k' l' m' \vert k+1 l+1 m-2>$. 
\\ \indent
The expectation value of this operator in the ground state of $H_g$ is 
represented as a function of $n$ and $g$ in Fig. 3(b).   
The results indicate that when $g$ is small $\Delta >0$, 
$<g \vert S_z^A S_z^B \vert g>=0$, and the main components of the ground 
state are the states $\vert k l m>$ with $k+l  > m$. 
However, when $g$ increases above $\sim 0.8$ 
the structure of the ground state has a sharp transition, and
the states $ \vert k l m>$ with $m > k+l$ become dominant.  
During this transition $\Delta$ vanishes (Fig. 3 (a)) and 
$<g \vert S_z^A S_z^B \vert g>$ becomes negative (Fig. 3 (b)), indicating 
the onset of the AF ground-state correlations. In the transition region the 
excitation energy 
$E_x$ has maxima when the number of pairs is even, and minima when it is odd 
(Fig. 2(b)).  This behavior is correlated with the change in the structure of 
the first excited state, whose main components can be obtained by
turning symmetrically one bilayer pair of the ground state into a pair 
within the layers. 
\\[.5cm] 
{\bf  IV. Summary and conclusions} \\[.5cm] \indent
The interlayer coupling in the cuprate compounds has a fundamental role
in the physics of the high-T$_c$ superconductivity, and in the recent
years an increasing experimental and theoretical effort was devoted for
understanding its mechanism \cite{and}. In this work the low-lying properties 
of two superconducting layers in interaction have been described considering 
two different coupling terms, one represented by the standard Josephson 
interaction, and one new, which is a "superexchange" pairing force between 
bilayer pairs. 
\\ \indent
The Josephson coupling between two superconductors is usually presented 
as a perturbation allowing the flow of the Josephson tunneling current 
\cite{kittel}. Complementary to this result, a TDMF \cite{mg} or RPA \cite{gi} 
calculation indicates that with this interaction the difference between the 
BCS phase angles of the two superconductors is not a constant, but oscillates 
in time, producing an excitation which could be observed as an antisymmetric 
low-lying resonance. These two pictures have been compared (Fig. 1) with the 
exact results obtained numerically for a restricted BCS Hamiltonian. 
\\ \indent
It was shown that the tunneling approach describes very well the energy of 
the first excited state when the number of pairs is odd and the strength of 
the Josephson coupling is small. In this case the bare ground state is double 
degenerate. The Josephson interaction rises this degeneracy, splitting the 
doublet into a non-degenerate ground state, symmetric between the two 
components, and the first excited state, which is antisymmetric. The two
components of the bare ground state become non-stationary states, interchanged 
periodically in time by quantum coherence oscillations. These oscillations 
can be pictured as a periodic exchange of a pair between the two layers. 
One should note, however, that at the relatively high temperature of the 
cuprate superconductors they might be completely suppressed by thermal 
decoherence \cite{mg3}.  \\ \indent
At moderate-to-strong coupling the excitation energy of the first level is 
very well described as a collective out-of-phase oscillation 
of the BCS gauge angles of the two layers. By contrast to the phase 
of a quantum wave function, the BCS angle is a collective variable 
defined only in the superconducting state, canonically conjugate to the 
particle number \cite{mg, mg84}.  Assuming a ratio $G_J/G = 0.5$ between 
the strength of the Josephson coupling $G_J$ and the pairing force within 
the layers $G$, the excitation energy of this resonance in YBa$_2$Cu$_3$O$_7$ 
is $\sim 50$ meV. This is a promising result, because indicates that by a 
more accurate estimate of $G_J/G$, the antisymmetric angular oscillation 
in the BCS gauge space might explain the 41 meV resonance observed in the 
neutron scattering experiments \cite{nse1, nse2}. 
\\ \indent
The effect of the Josephson coupling on the two-particle correlations 
was studied by calculating the pair transfer matrix element $\Delta$ 
between ground states. An increase of $\Delta$ indicates that the 
structure of the ground state is less sensitive to the addition or removal
of a pair, and becomes closer to the one of the BCS condensate. 
The results presented in Fig. 1 (b) and (d) show that the Josephson
coupling can induce the configuration mixing required for the 
occurrence of the superconductivity within the layers. At moderate-to-strong
coupling $\Delta$ is very close to the BCS gap (Fig. 1 (d)), indicating 
that the ground state of the whole system can be well approximated by a 
product of two independent BCS functions, one for each plane. 
\\ \indent 
The decomposition of the most general pairing interaction in a bilayer
system leads naturally to the Josephson coupling term, but also to a pairing 
interaction between "bilayer pairs", similar to the superexchange force
discussed in ref. \cite{and}. These pairs are formed by electrons 
from different planes with an antiparallel orientation of the spins, and 
therefore are related to the AF arrangement. The three different pair 
creation operators, (two for pairs within the layers and one for the bilayer 
pairs), together 
with their Hermitian conjugates generate by commutation the so(5) algebra. 
This result is important, because it shows that for a wide class of pairing 
Hamiltonians the many-body Hilbert space can be split in a series of 
finite-dimensional spaces of so(5) irreducible representation. In particular,
one can expect to find the ground state within a space ${\cal H}_\Omega$, 
carrier of a symmetric irrep of so(5). The construction of these spaces is a 
non-trivial problem in the Lie algebra representation theory, and it was 
presented in Sect. III. \\ \indent
The low-lying properties of two superconducting layers  coupled by
Josephson and bilayer pair interactions have been studied as a function of
the occupation fraction in Sect. IV. For the explicit realization was
used a set of so(5)$_r$ operators, but this choice is irrelevant as long as
the Hamiltonian contains only a single type of bilayer pairs ($R$ or $S$),
and is invariant to the so(5)$_r \leftrightarrow $so(5)$_s$ transformation.
The results indicate that when the strength of the bilayer pair interaction 
$g$ is $\sim 0.8 G$ or greater, 
there is a strong odd-even effect in the excitation energy of the first 
state (Fig. 2(b)), and the superconductivity within the planes disappears 
(Fig. 3 (a)). This transition is accompanied also by the onset of an AF 
arrangement between the spins of the two layers (Fig. 3 (b)), and is
practically independent of the occupation fraction. Therefore, within this
model an AF-SC transition at the variation of $n$ can be explained only by 
assuming that the ratio $g/G$ depends on $n$ and reaches the maximum at 
half-filling. The increase of the pairing interaction strengths within the 
planes at half-filling was noticed in ref. \cite{sisi}. The study of the 
concentration dependence of the interlayer coupling strengths appears 
therefore as a subject worth of further investigation. 
\\[2cm]
{\bf Acknowledgements} \\
The hospitality of Prof. M. Singh and of the Department of Physics and 
Astronomy, University of Western Ontario, is gratefully acknowledged.
Thanks are due to Prof. D. J. Rowe, University of Toronto, for many
stimulating discussions. \\[1cm]

\vskip1cm
\centerline{Table 1. The so(5) generators with negative roots}
\vskip0.5cm
\begin{tabular}{|c|c|c|c|c|}
\hline
so(5)   &  $f_{21} $     & $f_{20}$               & $f_{2 -1}$    & $f_{10}$    
\\ \hline
so(5)$_r$ & $P^\dagger_A$ & $R^\dagger / \sqrt{2}$ & $P^\dagger_B$ & 
$T_+ / \sqrt{2}$  \\ \hline
so(5)$_s$ & $-P^\dagger_A$ & $S^\dagger / \sqrt{2}$ & $P^\dagger_B$ & 
$\tau_+/ \sqrt{2}$  \\ \hline
\end{tabular}
\vskip1cm
{\bf Figure Captions} \\[.5cm]
Fig. 1. The excitation energy $E_x=E_1-E_0$ (a,c) and the pair transfer 
matrix element  $\Delta = G < g(N_p) \vert P^\dagger_A \vert g(N_p -1) >$ 
(b,d) as a function of the occupation fraction $n=N_p/2 \Omega$ when
$\Omega=12$, $G_A=G_B=1$, $g=0$ (dots joined by dashed lines).  
For comparison, $E_r$, $E_t$ and $\Delta_0 $ are represented by solid lines. 
Two values of the Josephson coupling strength are considered, 
$G_J=0.005$ (a),(b) and $G_J=0.2$ (c),(d). 
\\ 

Fig. 2. The ground state energy $E_0$ (a) and the excitation energy 
$E_x=E_1-E_0$ (b) as a function of the occupation fraction $n$ and the 
strength of the bilayer pairs interaction $g$ when 
$\Omega=6$, $G_A=G_B=1$ and $G_J=0.5$. 
\\

Fig. 3. The pair transfer matrix element $\Delta$ (a) and the correlation 
function  $<g (N_p) \vert S_z^A S_z^B \vert g (N_p) >$ (b) as a function of 
the occupation fraction $n$ and the strength of the bilayer pairs interaction 
$g$ when $\Omega=6$, $G_A=G_B=1$ and $G_J=0.5$.

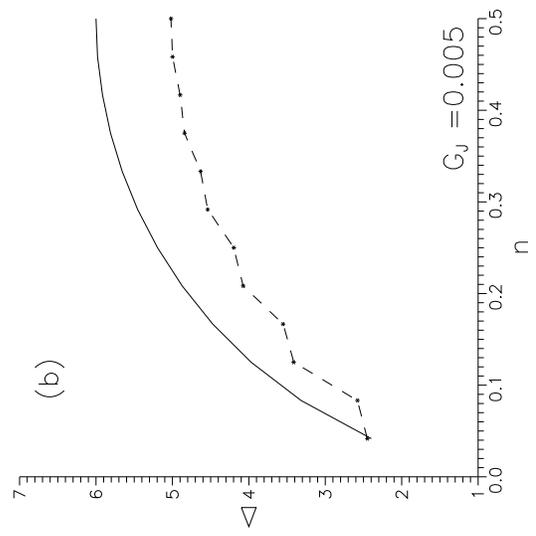
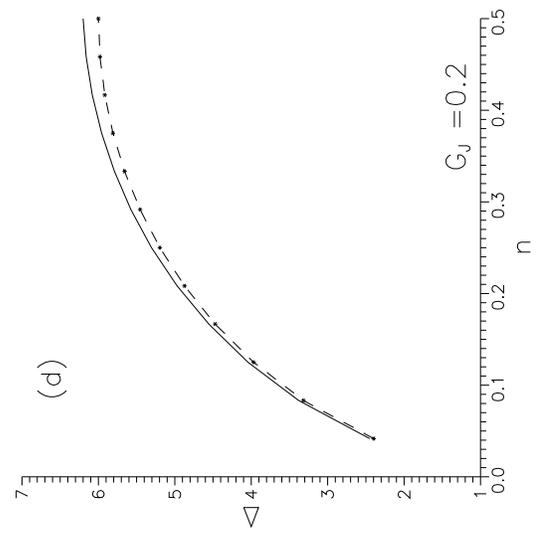
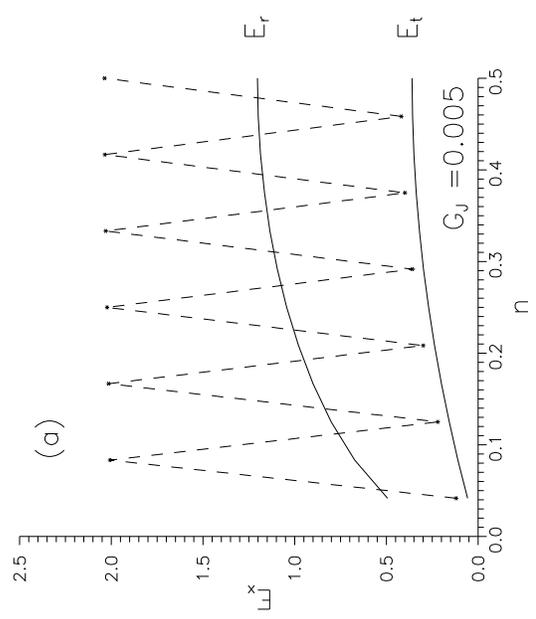
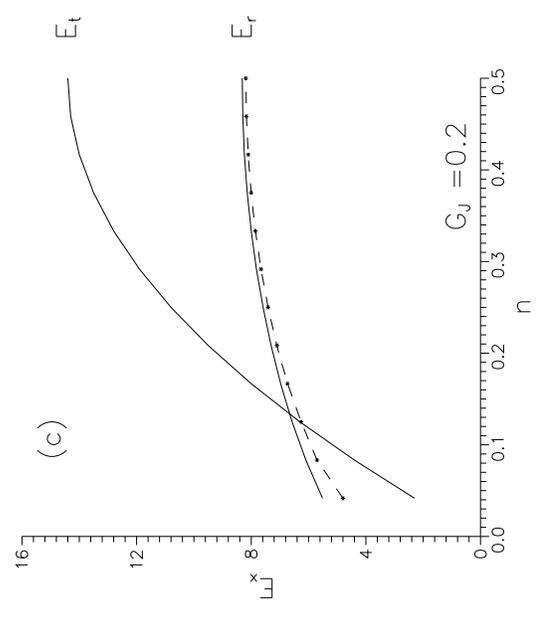
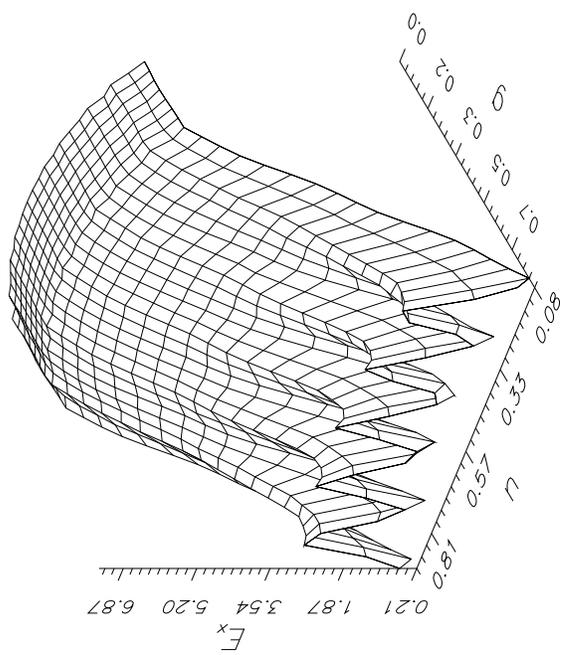
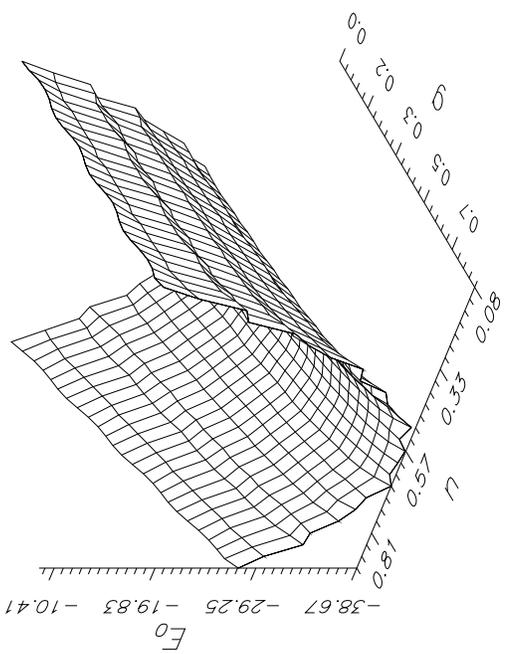
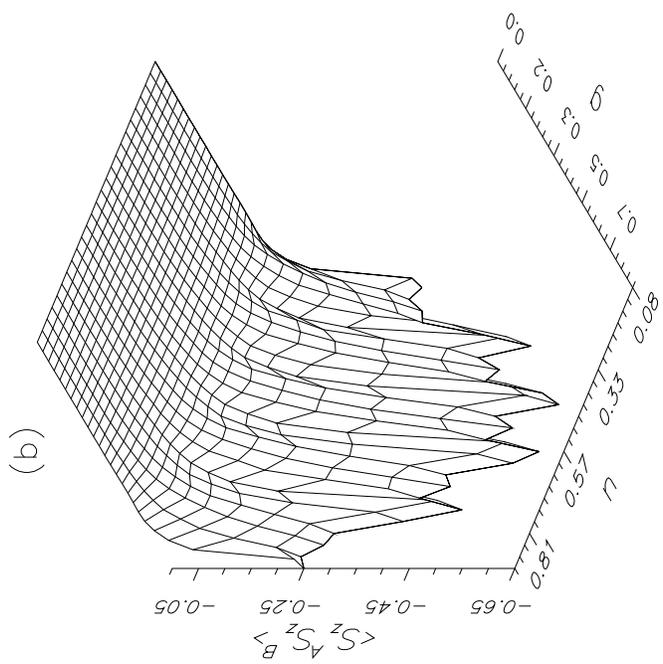
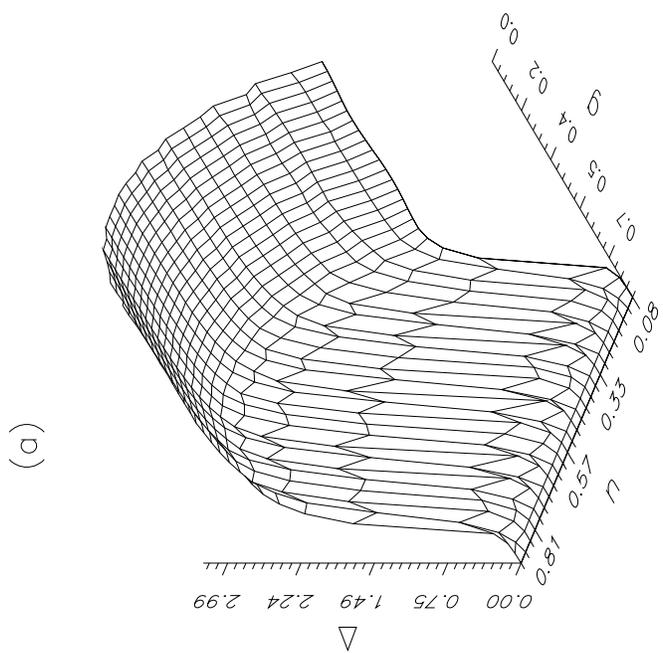
\end{document}